\documentstyle[aps,prl,multicol]{revtex}
\input psfig

\newcounter{saveeqn}
\newcommand{\alphaeqn}{\setcounter{saveeqn}{\value{equation}}%
\stepcounter{saveeqn}\setcounter{equation}{0}%
\renewcommand{\theequation}{\mbox{\arabic{saveeqn}\alph{equation}}}}
\newcommand{\reseteqn}{\setcounter{equation}{\value{saveeqn}}%
\renewcommand{\theequation}{\arabic{equation}}}

\begin{document}

\title{Entropy and dynamic properties of water below the \\ homogeneous
nucleation temperature}

\author{Francis~W. Starr$^{1}$, C.~Austen Angell$^{2}$, Robin
J. Speedy$^{3}$, and H.~Eugene Stanley$^{1}$}

\address{$^1$Center for Polymer Studies, Center for Computational
Science, and Department of Physics, \\ Boston University, Boston, MA
02215 USA}

\address{$^2$Department of Chemistry, Arizona State University, Tempe,
AZ, 85287, USA}

\address{$^3$7 Lindale Place, Waikanae Beach, Kapiti, New Zealand}

\date{July 20, 1999}
 
\maketitle

\begin{abstract}
Controversy exists regarding the possible existence of a transition
between the liquid and glassy states of water.  Here we use experimental
measurements of the entropy, specific heat, and enthalpy of both liquid
and glassy water to construct thermodynamically-plausible forms of the
entropy in the difficult-to-probe region between 150~K and 236~K.  We
assume there is no discontinuity in the entropy of the liquid in this
temperature range, and use the Adam-Gibbs theory -- which relates
configurational entropy to dynamic behavior -- to predict that dynamic
quantities such as the diffusion constant and the viscosity pass through
an inflection where the liquid behavior changes from that of an
extremely ``fragile'' liquid to that of a ``strong'' liquid.
\end{abstract}
\bigskip

\begin{multicols}{2}

It has been evident for at least three decades that there is a problem
connecting the thermodynamic behavior of water at normal temperatures to
that of glassy water (which is found below a glass transition
temperature $T_g \approx 136$~K)~\cite{debenedetti,angell-sare}.  Some
contributions assert that the liquid and glassy states should be
thermodynamically continuous, based on measurements of, e.g., specific
heat, entropy, and relaxation
times~\cite{combine1,ast73,jhm,jfhm94,sdshk}.  Other contributions argue
that the liquid and glass should be thermodynamically
distinct~\cite{angell-sare,spinodal,combine2,waterII}.  In particular,
refs.~\cite{jfhm94,waterII} focus on a thermodynamically-plausible form
for the entropy, and determine the limits on the entropy of the glass
that are consistent with the possibility of continuity.  The entropy of
the glass was subsequently measured~\cite{sdshk}, and found to be
consistent with (but does not require) thermodynamic continuity between
the liquid and glassy states.

As a simple illustration of the utility of thermodynamics to identify
the existence of a transition or other anomalous behavior of
thermodynamic properties, suppose we examine the experimental data for
the enthalpy $H$, entropy $S$, and specific heat $C_P$ of liquid water
at, e.g., $10^\circ$C and of ice Ih at, e.g., $-10^\circ$C.  The only
way to reconcile the large differences in $H$ and $S$ is to hypothesize
a discontinuity in $S$ or a large ``spike'' in $C_P$ in this temperature
range, even without knowledge that a first-order melting transition
occurs.  In other words using only thermodynamic data, one can place
relatively stringent limits on the thermodynamic behavior near the
melting transition.  Inspired by this fact, we perform test for the
presence of a ``dramatic change'' of the liquid thermodynamics below the
homogeneous nucleation temperature $T_H$ of supercooled water and above
the crystallization temperature $T_X$ of glassy water.  Specifically, we
use experimental data on the specific heat, entropy, and enthalpy in
both the liquid and glassy states to construct two
thermodynamically-plausible forms of the entropy in the
difficult-to-probe region between $T_X \equiv 150$~K and $T_H \equiv
236$~K at 1 atm~\cite{accessibility}.  We then use these forms for the
entropy, in conjunction with the theory of Adam and
Gibbs~\cite{AGtheory}, to predict behavior of the diffusion constant and
the viscosity.

To determine a reasonable form for the entropy $S = S(T,P)$, we first
focus several of the thermodynamic properties that facilitate
calculation of $S$ in the {\it experimentally-accessible\/} region and
also place strict limits on the possible behavior of $S$ in the {\it
difficult-to-probe} region $T_X<T<T_H$.  We define the excess enthalpy
$H_{\mbox{\scriptsize ex}} \equiv H_{\mbox{\scriptsize liquid}} -
H_{\mbox{\scriptsize crystal}}$, the excess entropy
$S_{\mbox{\scriptsize ex}} \equiv S_{\mbox{\scriptsize liquid}} -
S_{\mbox{\scriptsize crystal}}$, the difference of the liquid and
crystal entropies, and the excess specific heat

\begin{equation}
C_P^{\mbox{\scriptsize ex}} \equiv C_P^{\mbox{\scriptsize liquid}} -
C_P^{\mbox{\scriptsize crystal}} = T \left (\frac{\partial
S_{\mbox{\scriptsize ex}}}{\partial T} \right)_P.
\label{eq:cp}
\end{equation}

\noindent
Each of these quantities is known outside the difficult-to-probe region,
and in particular at the bounds $T_X$ and $T_H$ of the experimentally
difficult-to-probe region (Table~\ref{thermodynamics-table}).

$\bullet~T>T_H$: $H_{\mbox{\scriptsize ex}}(T_H) = H_{\mbox{\scriptsize
liquid}}(T_H) - H_{\mbox{\scriptsize crystal}}(T_H)$ has been measured
from the heat of crystallization of supercooled water~\cite{aos82}.  We
can relate measured values of $C_P^{\mbox{\scriptsize ex}}$ to
$S_{\mbox{\scriptsize ex}}$ by integrating Eq.~(\ref{eq:cp}),

\begin{equation}
S_{\mbox{\scriptsize ex}}(T) = S_{\mbox{\scriptsize
ex}}(T_{\mbox{\scriptsize M}}) - \int_T^{T_{\mbox{\scriptsize M}}}
\frac{C_P^{\mbox{\scriptsize ex}}}{T} dT \hspace{2cm}
[T<T_{\mbox{\scriptsize M}}]
\label{eq:Sex}
\end{equation}

\noindent
where $S_{\mbox{\scriptsize ex}}(T_{\mbox{\scriptsize M}}) = \Delta
S_{\mbox{\scriptsize F}}$, the entropy of fusion.  We numerically
evaluate the integral in Eq.~(\ref{eq:Sex}) for $T>T_H$, since we know
$C_P^{\mbox{\scriptsize liquid}}$ from recent bulk sample studies at
temperatures from $T_{\mbox{\scriptsize M}}$ down to
$-29^\circ$C~\cite{tfs-inpress}, (and by emulsion studies down to
$-37^\circ$C~\cite{aos82}), and we know $C_P^{\mbox{\scriptsize
crystal}}$ for all $T<T_{\mbox{\scriptsize M}}$~\cite{gs36}.

$\bullet~T<T_X$: $H_{\mbox{\scriptsize ex}}(T_X) = H_{\mbox{\scriptsize
liquid}}(T_X) - H_{\mbox{\scriptsize crystal}}(T_X)$ has been measured
from the heat of crystallization of glassy water~\cite{enthalpy}.
$C_P^{\mbox{\scriptsize ex}}$ below $T_X$ is known to be very small, and
may be taken to be nearly $T$-independent for $T\approx T_X$.
$S_{\mbox{\scriptsize ex}}(T_X)$ is known from the vapor pressure
experiments on the glass and and the crystal states~\cite{sdshk}.

$\bullet~T_X<T<T_H$: We construct two possible forms for
$S_{\mbox{\scriptsize ex}}$ for $T_X<T<T_H$ similar to the methods of
refs.~\cite{jfhm94,waterII}, but we now include the known value of
$S_{\mbox{\scriptsize ex}}(T_X)$.  $S_{\mbox{\scriptsize ex}}$ and
$C_P^{\mbox{\scriptsize ex}}$ fix the endpoints and the slopes of
$S_{\mbox{\scriptsize ex}}$ at $T_X$ and $T_H$, while the identity

\begin{equation}
H_{\mbox{\scriptsize ex}}(T_H) - H_{\mbox{\scriptsize ex}}(T_X) =
\int_{T_X}^{T_H} C_P^{\mbox{\scriptsize ex}} dT = 2910 \pm
30~\mbox{J/mol}
\label{enthalpy-eqn1}
\end{equation}

\noindent 
constrains the area bounded by $C_P^{\mbox{\scriptsize ex}}$ -- and so
also constrains the area bounded by $S_{\mbox{\scriptsize ex}}$, due to
Eq.~(\ref{eq:cp}).

Using these five thermodynamic constraints~\cite{constraints}, we
construct two possible forms of $S_{\mbox{\scriptsize ex}}$
[Fig.~\ref{s+c_p-fig}(a)] and the corresponding $C_P^{\mbox{\scriptsize
ex}}$ [Fig.~\ref{s+c_p-fig}(b)].  Curve 1 shows the case of continuity
with no transition, while curve 2 shows continuity with a
previously-discussed $\lambda$-transition~\cite{ast73,speedy87}.  We
obtain curve 1 in an {\it ad-hoc} fashion that satisfies the
thermodynamic constraints.  We use a closed form for
$C_P^{\mbox{\scriptsize ex}}$ and $S_{\mbox{\scriptsize ex}}$ in the
$\lambda$-transition case, assuming mean-field behavior~\cite{spinodal}

\alphaeqn
\begin{equation}
C_P^{\mbox{\scriptsize ex}}(T) = \left\{\begin{array}{l@{\quad~~\quad}l}
  a + \frac{b_+}{\sqrt{T/T_\lambda-1}} & T>T_\lambda \\
  a + \frac{b_-}{\sqrt{1-T/T_\lambda}} & T<T_\lambda \end{array} \right. .
\label{eq:cP-mf}
\end{equation}

\noindent from which we obtain

\begin{equation}
S_{\mbox{\scriptsize ex}}(T) = \left\{ \begin{array}{l@{\quad~~\quad}l}
	S_0 + a \log{T} + 2b_+ \tan^{-1}{\sqrt{T/T_\lambda-1}} &
	T>T_\lambda \\ S_0 + a \log{T} - 2b_-
	\tanh^{-1}{\sqrt{1-T/T_\lambda}} & T<T_\lambda \end{array}
	\right. .
\end{equation}
\reseteqn

\noindent In principle, the five free parameters ($S_0$, $a$, $b_\pm$,
and $T_\lambda$) may be determined by the five thermodynamic
constraints.  However, such a procedure yields an experimentally
unreasonable value of $T_\lambda = 269$~K.  To obtain reasonable values
for the parameters, we choose $S_{\mbox{\scriptsize ex}}(200~\mbox{K}) =
3.4$~J/(K$\cdot$mol), $C_P^{\mbox{\scriptsize ex}}(200~\mbox{K}) =
15$~J/(K$\cdot$mol), and $T_\lambda = 225$~K.  This fixes the remaining
free parameters in Eq.~(4)~\cite{lambda}.

Curves 1 and 2 for $S_{\mbox{\scriptsize ex}}$ and
$C_P^{\mbox{\scriptsize ex}}$ both show sharp changes in their behavior
just below 230~K.  Note that a significantly less sharp change in
$S_{\mbox{\scriptsize ex}}$ than shown would not satisfy the constraint
of Eq.~(\ref{enthalpy-eqn1}), as can also be seen by rewriting
constraint (\ref{enthalpy-eqn1}) in terms of the area bounded by
$S_{\mbox{\scriptsize ex}}$.  From integration by-parts, we find

\begin{equation}
\int_{T_X}^{T_H} S_{\mbox{\scriptsize ex}} dT =
\Big[TS_{\mbox{\scriptsize ex}} - H_{\mbox{\scriptsize ex}}
\Big]_{T_X}^{T_H} = 422 \pm 30~\mbox{J/mol}.
\label{enthalpy-eqn2}
\end{equation}

\noindent 
A more gradual change in $S_{\mbox{\scriptsize ex}}$ below $T_H$ than
shown in Fig.~\ref{s+c_p-fig}(a) would bound an area $\int_{T_X}^{T_H}
S_{\mbox{\scriptsize ex}} dT$ {\it larger} than 422~J/mol.  Furthermore,
the inflection in $S_{\mbox{\scriptsize ex}}$ [Fig.~\ref{s+c_p-fig}]
must occur at $T \gtrsim 215$~K, as moving the inflection to a
significantly lower temperature would also yield an area too
large.

We do not hypothesize a discontinuous form for the entropy because the
magnitude of the possible discontinuity is unknown.  While the data in
Table~\ref{thermodynamics-table} and the thermodynamic constraints do
not require a discontinuity in $S_{\mbox{\scriptsize conf}}$ below
$T_H$, the data can not rule out the possibility.  Furthermore, behavior
of $C_P^{\mbox{\scriptsize ex}}$ resembling a step-function is also
possible.  However, the accelerating increase of $C_P^{\mbox{\scriptsize
ex}}$ approaching the inaccessible region from above makes the
$\lambda$-transition a more natural choice.  On the other hand, we
emphasize that using the available data, it is impossible to distinguish
the correct from of $C_P^{\mbox{\scriptsize ex}}$, or the exact location
of any possible transition (only that an anomaly occurs within the
approximate range of $200~K<T<230~K$).

We now consider the possible implications of the approximate forms for
$S_{\mbox{\scriptsize ex}}$ on the dynamic behavior below $T_H$.  The
entropy-based Adam-Gibbs theory~\cite{AGtheory} has been used to
describe the relaxation of liquids approaching their glass
transitions~\cite{AGsuccess}, and provides an explanation for the
variation of diffusion constant $D$ and viscosity $\eta$ (or other
characteristic dynamic quantity) in the anomalous range below
$-20^\circ$C~\cite{afwb76}.  We use the prediction

\begin{equation}
\eta = \eta_0 \exp\left(\frac{A}{T S_{\mbox{\scriptsize conf}}}\right).
\label{AG-eqn}
\end{equation}

\noindent 
Here $A$ is a constant~\cite{VFTnote}.  The configurational entropy of
the liquid, $S_{\mbox{\scriptsize conf}} \equiv S_{\mbox{\scriptsize
liquid}} - S_{\mbox{\scriptsize vib}}$, can be understood as the entropy
attributable to the various basins the liquid can sample in the energy
landscape picture~\cite{debenedetti,glasses95}.  The vibrational
component $S_{\mbox{\scriptsize vib}}$ of the entropy is attributable to
the thermal excitation the liquid experiences in the basin sampled, so
we may approximate $S_{\mbox{\scriptsize vib}}^{\mbox{\scriptsize
liquid}} \approx S_{\mbox{\scriptsize vib}}^{\mbox{\scriptsize
crystal}}$.  For typical crystals, $S_{\mbox{\scriptsize conf}} \approx
0$, since the crystal samples a negligible number of basins.  Hence
$S_{\mbox{\scriptsize crystal}} \approx S_{\mbox{\scriptsize vib}}$,
from which it follows that for the liquid $S_{\mbox{\scriptsize conf}}
\approx S_{\mbox{\scriptsize ex}}$.  The approximation
$S_{\mbox{\scriptsize vib}}^{\mbox{\scriptsize liquid}} \approx
S_{\mbox{\scriptsize vib}}^{\mbox{\scriptsize crystal}}$ is quite good
for liquids near $T_g$~\cite{goldstein}, as the liquid typically samples
only the deepest basins in the energy landscape, with small harmonic
excitations about the minimum, similar to the crystal behavior.  At
higher $T$, the liquid explores a much greater region of the landscape,
and is no longer localized in a single basin for an extended time.  Thus
at higher $T$, the approximation that $S_{\mbox{\scriptsize
vib}}^{\mbox{\scriptsize liquid}} \approx S_{\mbox{\scriptsize
vib}}^{\mbox{\scriptsize crystal}}$ may not work as
well~\cite{mct-note}.

In the case of ice, there exists a residual entropy
$S_{\mbox{\scriptsize res}}$ at zero temperature due to proton
disorder~\cite{pauling35}.  When we subtract $S_{\mbox{\scriptsize
crystal}}$ from $S_{\mbox{\scriptsize liquid}}$, we also implicitly
subtract the contribution arising from $S_{\mbox{\scriptsize res}}$,
which should also contribute to the available configurations of the
liquid, so we include $S_{\mbox{\scriptsize res}}$
explicitly
\begin{equation}
S_{\mbox{\scriptsize conf}} = S_{\mbox{\scriptsize ex}} +
S_{\mbox{\scriptsize res}}.
\label{conf-entropy-eqn}
\end{equation}

We use this form of $S_{\mbox{\scriptsize conf}}$ to predict the
behavior of $\eta$ and $D$~\cite{hedge} for $T \le T_H$.  We select
parameters~\cite{parameters} in Eq.~(\ref{AG-eqn}) to fit the
experimental values of $D$~\cite{gillen} and $\eta$~\cite{viscosity}
[Fig.~\ref{AGpredictions-fig}].  The non-Arrhenius behavior for $T
\gtrsim 230$~K is typical for a fragile liquid~\cite{glasses95}.  As $T
\rightarrow T_g$, we find $\eta(T_g) \approx 10^{16}$~Poise.  This is
roughly 3 orders of magnitude larger than $\eta(T_g)$ expected from
experiments~\cite{johari98}, but is not unreasonable considering that a
number of approximations that were necessary, and further that we have
extrapolated over 14 orders of magnitude from experimental data covering
only one order of magnitude.  Furthermore, excluding
$S_{\mbox{\scriptsize res}}$ in Eq.~(\ref{conf-entropy-eqn}), results in
an absurd value $\eta(T_g) \approx 10^{43}$ Poise, emphasizing the
importance of including $S_{\mbox {\scriptsize res}}$ as part of
$S_{\mbox {\scriptsize conf}}$~\cite{moynihan,fs-comment}.

The dramatic change in entropy around 225~K required by the physical
constraints is reflected by the ``kinks'' that appear in $D$ and $\eta$
at $T \approx 225$~K.  In contrast to the fragile behavior for $T
\gtrsim 220$~K, the behavior for $T \lesssim 220$ is characteristic of a
strong liquid~\cite{glasses95} -- i.e. Arrhenius behavior with an
appropriate activation energy.  We find $E_D/RT_g \approx 28$,
$E_{\eta}/RT_g \approx 29$, larger than the expected activation energy
$E/RT_g \approx 14$ for an ``ideal'' strong liquid~\cite{glasses95}, but
much less than that of a fragile liquid ($E/RT_g \approx 80-100$).
Furthermore, adjustment of the constants in Eq.~(\ref{AG-eqn}) such that
$\eta(T_g) = 10^{13}$~Poise, as expected from
experiments~\cite{johari98}, yields a value of $E_{\eta}/RT_g$ much
closer to the ``ideal'' value for a strong liquid.  These results
support the hypothesis that the fragile behavior of water shows a change
to strong behavior for $T \lesssim 220$~K.  Water is also expected to be
a strong liquid near $T_g \approx 136$~K, based on measurements of the
change in $C_P$ near $T_g$, and the width of the glass
transition~\cite{angell93}.  Such a crossover from fragile to strong
behavior is not typical of liquids, and merits further experimental
scrutiny~\cite{comment}.

We wish to thank B.D.~Kay, S.~Sastry, F.~Sciortino, R.S.~Smith, and
M.~Yamada for enlightening discussions.  We especially thank
C.T. Moynihan for his important contributions.  FWS is supported by a
NSF graduate fellowship.  CAA acknowledges support form a NSF Solid
State Chemistry grant DMR-9108028-002.  RJS acknowledges support from
the Marsden Fund through contract GRN 501.  The Center for Polymer
Studies is supported by NSF grant CH9728854.

\begin{table}
\caption{Thermodynamic properties of water at 1 atm at 150~K and 236~K.
Here, $X_{\mbox{\scriptsize ex}} \equiv X_{\mbox{\scriptsize liquid}} -
X_{\mbox{\scriptsize crystal}}$, the excess quantity $X$ of the liquid
value relative to the ice Ih value.  The uncertainties of
$S_{\mbox{\scriptsize ex}}$ and $H_{\mbox{\scriptsize ex}}$ are taken
from~\protect\cite{speedy87}, which contains arguments supporting the
reliability of the data.}
\medskip
\begin{tabular}{c|lc}
	&	$T_X = 150$~K	     &	$T_H = 236$~K \\
\tableline
$C_P^{\mbox{\scriptsize ex}}$ [J/(K$\cdot$mol)] & 
$\approx 2$~\cite{jhm,jfhm94,hk88} &
$69.2 \pm 0.5$~\cite{tfs-inpress,aos82,gs36} \\
$S_{\mbox{\scriptsize ex}}$ [J/(K$\cdot$mol)] & 
$1.7 \pm 1.7$~\cite{sdshk} &
$15.2 \pm 0.1$~\cite{speedy87,aos82,gs36} \\
$H_{\mbox{\scriptsize ex}}$ [J/mol] & 
$1380 \pm 20$~\cite{enthalpy} &
$4290 \pm 20$~\cite{speedy87,aos82} \\
\end{tabular}
\label{thermodynamics-table}
\end{table}

\newbox\figa
\setbox\figa=\psfig{figure=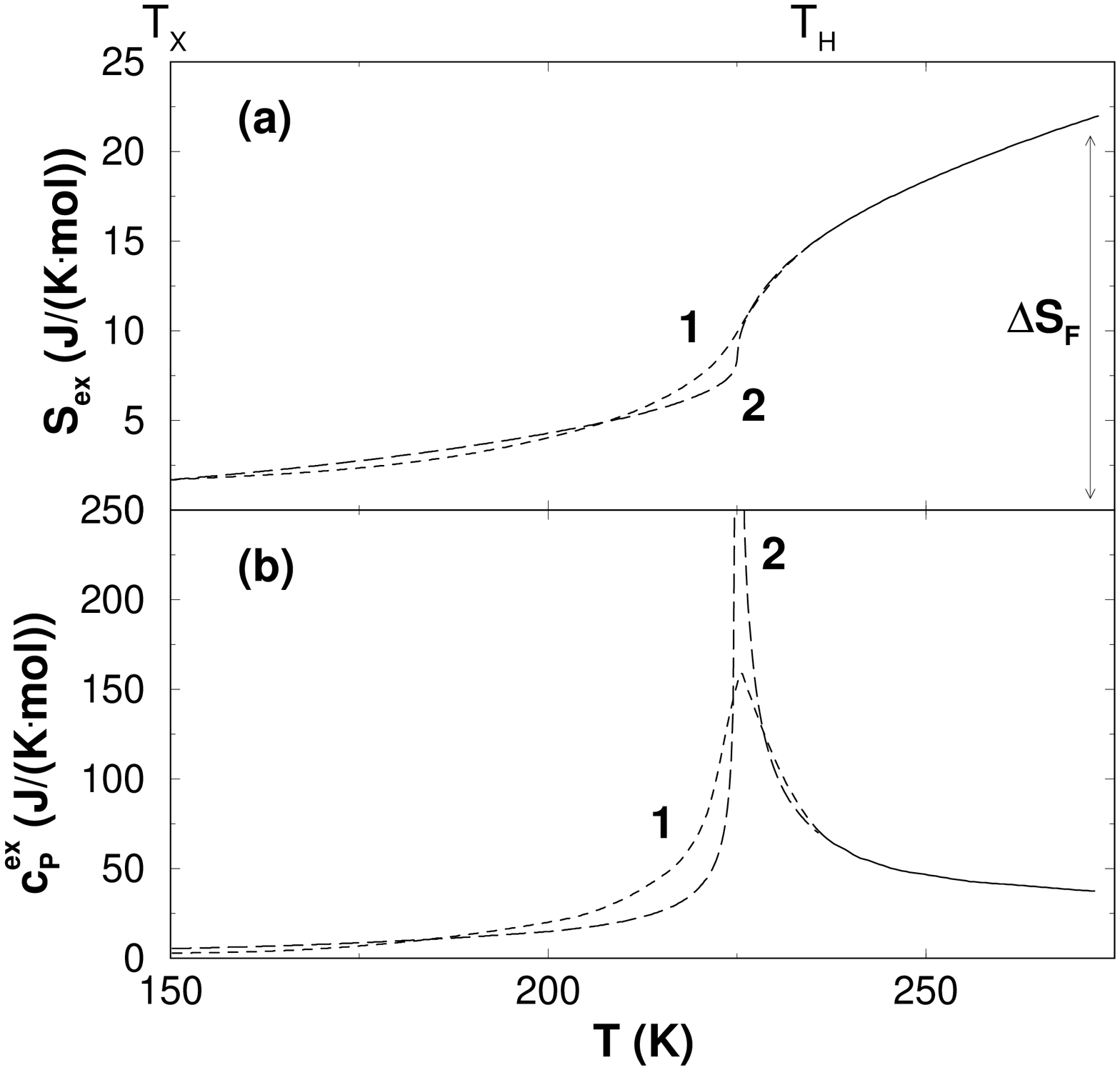,width=3.5in,angle=-90}
\begin{figure*}[htbp]
\begin{center}
\leavevmode
\centerline{\box\figa}
\narrowtext
\caption{ (a) Possible forms for the excess entropy
$S_{\mbox{\scriptsize ex}}$ in the experimentally inaccessible region.
Curve 1 corresponds to no transition, while curve 2 shows a
$\lambda$-transition at 225~K.  Any thermodynamically plausible form of
$S_{\mbox{\scriptsize ex}}$ (without a discontinuity) can vary only
slightly from these form (due to the uncertainty in
$H_{\mbox{\scriptsize ex}}$).  The entropy of fusion $\Delta
S_F=21.8$~J/(K$\cdot$mol) for freezing at 273~K is indicated by the
arrow.  (b) Constant pressure excess specific heat
$C_P^{\mbox{\scriptsize ex}} = T (dS_{\mbox{\scriptsize ex}}/dT)_P$ for
the possible forms of $S_{\mbox{\scriptsize ex}}$ shown in (a).  Curve 2
has a diverging $C_P^{\mbox{\scriptsize ex}}$ at the
$\lambda$-transition temperature.}
\label{s+c_p-fig}
\end{center}
\end{figure*}

\newbox\figa
\setbox\figa=\psfig{figure=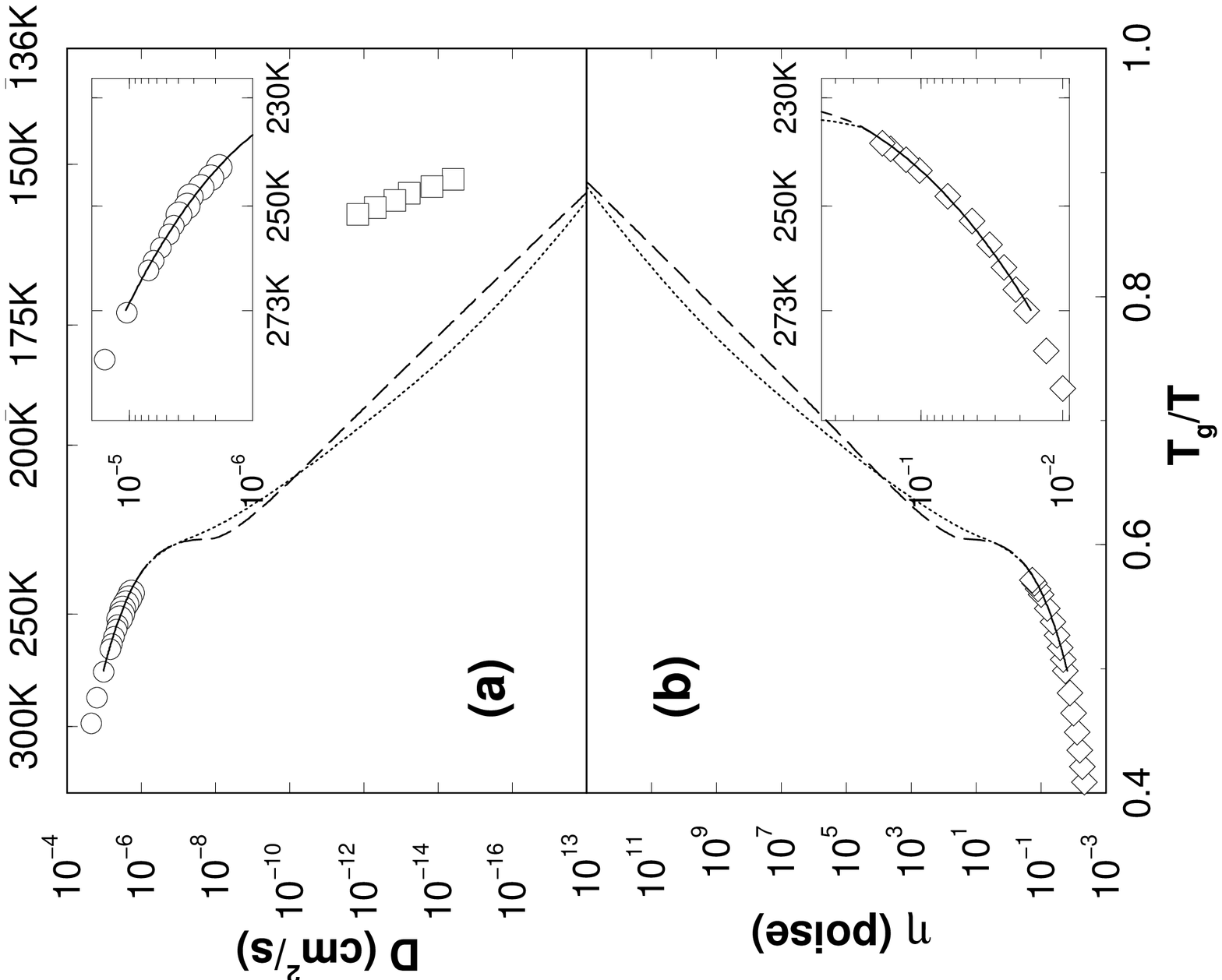,width=3.5in,angle=-90}
\begin{figure*}[htbp]
\begin{center}
\leavevmode
\centerline{\box\figa}
\narrowtext
\caption{ (a) Diffusivity predicted by Eq.~(\protect\ref{AG-eqn}).  The
experimental data ($\circ$) for $T > 235$K are
from~\protect\cite{gillen}.  The data for $T<160$~K ($\Box$) are
from~\protect\cite{smith-kay} (b) Fit of $S_{\mbox{\scriptsize ex}}$ to
viscosity using the same procedure.  Experimental data ($\diamond$) are
from \protect\cite{viscosity}.  Both (a) and (b) show behavior expected
for a strong liquid for $T \protect\lesssim 220$~K -- i.e. Arrhenius
behavior with an activation energy $\approx T_g/3$ (in units of
kJ/mol)~\protect\cite{glasses95}.  The insets show the quality of the
fit in the region where experimental data are available.}
\label{AGpredictions-fig}
\end{center}
\end{figure*}

\end{multicols}{2}


\begin{references}
\bibitem{constraints} The five constraints are the known values of
$S_{\mbox{\scriptsize ex}}$ and $C_P^{\mbox{\scriptsize ex}}$ at $T_X$
and $T_H$, plus the limitation on the area bounded by
$S_{\mbox{\scriptsize ex}}$ between $T_X$ and $T_H$.
\bibitem{lambda} The values we obtain for the $\lambda$-transition are:
$T_\lambda = 225$~K, $S_0 = 34.3$~J/(K$\cdot$mol), $a = -4.8$~J/mol,
$b_+ = 16.4$~J/mol, and $b_- = 6.6$~J/mol.  We allow for the
possibility that $b_+$ and $b_-$ differ because the prefactor on the
diverging term of Eq.~(\ref{eq:cP-mf}) typically depends on whether the
transition is approached from temperatures above or below $T_\lambda$.
\bibitem{VFTnote} The Vogel-Fulcher-Tamman form $\eta = \eta_0
\exp(B/(T-T_0))$ for the temperature dependence of viscosity and
characteristic times of liquids at low temperature can be obtained from
Eq.~(\ref{AG-eqn}) by assuming that $C_P^{\mbox{\scriptsize ex}} \propto
T^{-1}$.  Note that $T_0 < T_g$ is typically associated with an
underlying ``ideal'' glass transition.
\bibitem{parameters} 
For the diffusion, we use $D_0 = 5.55 \times 10^{-4}$~cm$^2$/s and $A =
-27.4$~kJ/mol.  For the viscosity, we use $\eta_0 = 1.99 \times
10^{-4}$~Poise and $e = 30.4$~kJ/mol.  These parameters were obtained by
fitting $S_{\mbox{\scriptsize conf}}$ to the experimental data in the
region between $T=235$~K and 273~K.  T
\bibitem{fs-comment} An alternative procedure is to fix $\eta(T_g)
\approx 10^{13}$~Poise, and determine the value of $S_{\mbox
{\scriptsize}}$ is consistent with this $\eta$.
\bibitem{comment} The expectation that water is strong near $T_g$ does
not agree with recent data~\cite{smith-kay}, perhaps indicating that the
Adam-Gibbs predictions are not valid for water.
\end{references}
\end{document}